\documentstyle[aps,psfig]{revtex}
\begin{document}
%

\title{{\Large\bf Effects of Magnetic Order on the Upper Critical Field of
UPt$_3$}}
\author{J.A. Sauls}
\address{Department of Physics \& Astronomy,
Northwestern University, Evanston, IL 60208 }
\date{December 13, 1994}

\maketitle

\begin{abstract}

\centerline{{\bf Abstract}}
{\small
\noindent

I present a Ginzburg-Landau theory for hexagonal oscillations of the
upper critical field of UPt$_3$ near $T_c$. The model is based on a
$2D$ representation for the superconducting order parameter,
$\vec{\eta}=(\eta_1,\eta_2)$, coupled to an in-plane AFM order parameter,
$\vec{m}_s$. Hexagonal anisotropy of $H_{c2}$ arises from the weak
in-plane anisotropy energy of the AFM state and the coupling of the
superconducting order parameter to the staggered field. The model
explains the important features of the observed hexagonal
anisotropy [N. Keller, {\it et al.}, Phys. Rev. Lett. {\bf
73}, 2364 (1994).] including: (i) the small magnitude, (ii) persistence
of the oscillations for $T\rightarrow T_c$, and (iii) the change in
sign of the oscillations for $T> T^{*}$ and $T< T^{*}$ (the temperature
at the tetracritical point). I also show that there is a
low-field crossover (observable only very near $T_c$) below which the
oscillations should vanish.
}

\end{abstract}


\section{Introduction}

The heavy Fermion superconductor UPt$_3$ has attracted a great deal of
attention because of its remarkable low temperature phase diagram.  The
key features are: (i) the existence of two superconducting phases in
zero field with a small separation of the transition temperatures,
$\Delta T_{c}\simeq 50\,mK$, compared to $T_{c}\simeq 500\,mK$, (ii)
the existence of three superconducting phases in a magnetic field (iii)
with phase boundaries that meet at a tetracritical point $(T^{*},
H^{*})$ on the upper critical field line.\cite{bru90,ade90}
Interpretations of this phase diagram fall into two main categories of
theoretical models: (i) theories based on two symmetry un-related order
parameters which are accidentally nearly degenerate,\cite{luk91,che93}
and (ii) models based on a single multi-component order parameter
belonging to a higher dimensional representation in which the
degeneracy is lifted by a weak symmetry breaking field
(SBF).\cite{hes89,mac89,mac91,joy91,luk94,sau94} UPt$_3$ is a
close-packed hexagonal crystal with two formula units per unit cell;
the point group is $D_{6h}$. Two candidates for an intrinsic SBF have
been identified experimentally. Neutron scattering experiments show
that UPt$_3$ develops an anti-ferromagnetic order parameter,
$\vec{m}_s\perp\hat{c}$, below a N\'eel temperature of $T_N\simeq 5$
K.\cite{aep88} The magnitude of the ordered moment is small, $\sim
0.02\mu_B$ per U atom, and is directed in the basal plane thus breaking
the in-plane hexagonal symmetry.  Evidence in support of an AFM
symmetry breaking field was found from pressure studies of the
superconductivity and AFM order. Heat capacity measurements show that
the splitting of the superconducting transitions is suppressed under a
hydrostatic pressure of $p_c\simeq 3.8$ kbar,\cite{hay92} while neutron
scattering measurements show the supression of AFM order at roughly the
same pressure, $p_{*}\simeq 3.2$ kbar.\cite{tra91} However, the
observed magnetic Bragg peaks indicate finite-range AFM order with a
correlation length $\xi_{afm}\simeq 250\,\AA$ depending on the crystal,
which has led to discussion of whether or not the AFM order is
intrinsic property of UPt$_3$.\cite{vor92} Alternatively, a macroscopic
strain field was proposed\cite{min93} as a possible SBF based on the
experimental observation of a complex incommensurate structural
modulation in UPt$_3$ with characteristic wavelengths of order 10-50
lattice spacings and typical domain sizes of order $10^4\,\AA$.
However, these structural modulations seem unlikely to be the principal
SBF responsible for the double transition since the high formation
temperature suggests that they persist to pressures well above that
required to destroy the double transition and the AFM
order.\cite{mid93}

In a recent article, Keller, {\it et al.}\cite{kel94} reported the
observation of a weak hexagonal modulation of the upper critical field
of UPt$_3$ as a function of the orientation of the field in the basal
plane. The key features of these measurements are: (i) the small
magnitude of the oscillations, $|\delta H_{c2}^{hex}/H_{c2}|\lesssim
0.02$, (ii) persistence of the oscillations for $T\rightarrow T_c$, and
(iii) the change in sign of the oscillations for $T> T^{*}$ and $T<
T^{*}$, where $T^{*}$ is the temperature at the tetracritical point. In
this paper I show that hexagonal oscillations such as these are a consequence
of the
in-plane AFM symmetry breaking field above $T_c$ that is responsible
for the double transition in UPt$_3$.

\section{Perfect Hexagonal Symmetry}

In order to sharpen the argument I briefly review the important theoretical
results on the anisotropy of the in-plane upper critical field in {\it
any} perfectly hexagonal superconductor. First consider a hexagonal
superconductor described by a one-component order parameter, $\eta$.
This case includes conventional superconductors for which $\eta$ is
invariant under $D_{6h}$, as well as unconventional cases where $\eta$
changes sign under one or more symmetry operations.  In the
Ginzburg-Landau (GL) limit the free energy functional for any
superconducting order parameter belonging to a one-dimensional representation
is
\begin{equation}
{\cal F} = \alpha(T)|\eta|^2 + \beta |\eta|^4
			+ \kappa_{||} (D_i\eta)(D_i\eta)^*
			+ \kappa_{\perp} (D_z\eta)(D_z\eta)^*
\,,
\end{equation}
where $D_i=\nabla_i+2e/\hbar c A_i$ and repeated indices are summed
over $(x,y)$. An important point here is that the GL functional has an
accidental symmetry; to second order in the gradients there are no
terms in the functional that differentiate hexagonal symmetry
($D_{6h}$) from cylindrical symmetry ($D_{\infty h}$). Consequently,
the upper critical field is independent of the orientation of $\vec{H}$
in the basal plane; $H_{c2}^{\perp}
=\frac{\hbar
c}{2e}|\alpha(T)|/\sqrt{\kappa_{||}\kappa_{\perp}}\sim(1-T/T_c)$.
Hexagonal anisotropy of $H_{c2}^{\perp}$ shows up only at lower
temperature when higher-order gradients become significant. The
lowest-order contribution to the free energy functional that is
invariant under $D_{6h}$, but not $D_{\infty h}$, is
\begin{equation}
{\cal F}_{hex} = \kappa_6\, |(D_x + i D_y)^{3}\eta|^2
\,.
\end{equation}
It is straight-forward to show that this term leads to hexagonal oscillations
of $H_{c2}^{\perp}(T)$ of the form,
\begin{equation}
\delta H_{c2}^{\perp}(\vartheta,T)\propto
	\kappa_6\,(1-T/T_c)^3
		\, * \, \cos(6\vartheta)
\,,
\end{equation}
to leading order in $\kappa_6$. The key point is that the hexagonal
anisotropy vanishes as $(1-T/T_c)^3$ as $T\rightarrow T_c$.
This result is not limited to
superconductors with a one-dimensional order parameter.

Burlachkov\cite{bur85} showed that $H_{c2}^{\perp}$ is isotropic for
fields in the basal plane, in the GL limit, for {\it any} of the
two-dimensional representations of $D_{6h}$. In this case the GL functional can
be written as
\begin{equation}\label{f_sc}
\begin{array}{l}
{\cal F}_{sc} = \alpha(T) |\eta_i|^2
			+ \beta_1 |\eta_i|^4
			+ \beta_2 |\eta_i^2|^2
			+ \kappa_4 (D_z\eta_i)(D_z\eta_i)^* \\ \hspace{4em}
			+ \kappa_1 (D_i\eta_j)(D_i\eta_j)^*
			+ \kappa_2 (D_i\eta_i)(D_j\eta_j)^*
 			+ \kappa_3 (D_i\eta_j)(D_j\eta_i)^*
\,,
\end{array}
\end{equation}
where the order parameter $(\eta_1,\eta_2)$ transforms according to one
of the 2D irreducible representations (E$_1$ or E$_2$) of $D_{6h}$. The
isotropy of $H_{c2}^{\perp}$ in the GL limit again follows because to
second-order in the gradients and in $(\eta_1,\eta_2)$ the
GL functional is invariant under the larger group, $D_{\infty h}$.
Again, one has to examine sixth-order terms (therefore,
of order $(1-T/T_c)^3$) in order to develop hexagonal anisotropy
in $H_{c2}^{\perp}$.\cite{vin94}

Recently, Mineev examined GL models based on two accidentally nearly
degenerate 1D representations.\cite{min94} Such a models have been
investigated by several authors\cite{luk91,che93} as candidate
theories of the H-T phase diagram of UPt$_3$. An important feature of
these models is that they do not rely on the coupling of the order
parameter to a SBF that reflects a weak breaking of hexagonal symmetry
above $T_c$.  The basic result for the anisotropy of $H_{c2}^{\perp}$
is the same; hexagonal oscillations vanish as $(1-T/T_c)^3$ near
$T_c$.\cite{min94} Thus, the experimental observation of hexagonal
oscillations of $H_{c2}^{\perp}$ in UPt$_3$ for $T\rightarrow T_c$ is
in conflict with any GL model of the superconducting phases that
is based on perfect hexagonal symmetry above $T_c$.

Below I show that
hexagonal oscillations of $H_{c2}^{\perp}(T)$ do appear in the GL limit
($T\approx T_c$) for the class of models based on a 2D superconducting
order parameter coupled to an AFM symmetry breaking field present above
$T_c$. It may initially seem odd that one obtains six-fold
oscillations from an AFM order parameter that {\it reduces} the
symmetry from hexagonal to orthorhombic. Six-fold oscillations of
$H_{c2}^{\perp}$ are shown to be a natural consequence of a small
in-plane anisotropy energy of the AFM order parameter, while
orthorhombic anisotropy of $H_{c2}$ appears in the limit of large
in-plane anisotropy.

\section{Broken Symmetry Model}

The free energy functional for a 2D superconducting order parameter,
$\vec{\eta}$, coupled to an AFM order parameter, $\vec{m}_s$, is the
sum of (i) ${\cal F}_{sc}$, which represents the GL functional for the
superconducting order parameter, (ii) ${\cal F}_{afm}$, which is the
free energy for the AFM phase above $T_c$, and (iii) ${\cal
F}_{sc-afm}$, which represents the coupling between superconductivity
and anti-ferromagnetism. The general form for the superconducting GL
functional in the absence of a SBF has been discussed by many
authors\cite{sig91}, and is given in eq. \ref{f_sc}. First, consider the
magnetic free energy above $T_c$.

\subsection{Free Energy of the AFM state above $T_c$}

For $T < T_N$ the mean-field approximation for the magnetic free
energy functional should be reliable provided the correct magnetic
ordering has been identified. I assume an order parameter,
$\vec{m}_s=(m_x, m_y, m_z)$, describing AFM correlations between
neighboring U atoms in a double unit cell, and start from a
Landau functional for the magnetic free energy density,
\begin{equation}\label{f_afm}
\begin{array}{l}
{\cal F}_{afm} = a(T)|\vec{m}_s|^2 + b|\vec{m}_s|^4 + ... \\ \hspace{5em}
	+ a_z m_z^2 + c'\,Re(m_x + i m_y)^6 + g (\vec{m}_s\cdot\vec{H})^2
\,.
\end{array}
\end{equation}
The first line of terms is invariant under the full spin-rotation
group. They are primarily due to exchange interactions in typical magnetic
materials. Thus, neglecting the anisotropy terms, one has $m_0 =
\sqrt{|a(T)|/2b}\propto|T-T_N|^{1/2}$ for the magnitude of the AFM
order parameter in the exchange approximation.

The third term represents the leading order uniaxial anisotropy energy.
Anisotropy energies arise from spin-orbit interactions and are
typically small compared to the exchange terms. I assume $a_z > 0$
which favors in-plane AFM order.

The sixth-order term in $\vec{m}_s$ is the leading term in a GL
expansion for the in-plane anistropy energy. The in-plane anisotropy
energy is assumed to be a small perturbation to the AFM exchange
energy, {\it i.e.}
\begin{equation}
U_{anis}=c'\,m_0^6\, \ll \, U_{exch} = \frac{1}{2}|a(T)| m_0^2
\,.
\end{equation}
In the absence of an external field the
in-plane anistropy energy leads to six degenerate minimum energy
orientations for $\vec{m}_s$, which are equivalent to three
commensurate wavevectors in a spin-density wave
description. There are two sets of
preferred orientations depending on the sign of the coefficient $c'$.
If $c' > 0$ the moments prefer the set of alignments, $\{\vartheta_n =
n\pi/3 + \pi/6; n=0, ..., 5\}$ that includes the $\vec{a}^*$ axis,
while for $c' <0 $ the set of minimum energy orientations are shifted
by $\pi/6$, and includes the $\vec{a}$ axis [I use the notation in Ref.
\onlinecite{kel94}].

The Zeeman energy is quadratic in $\vec{H}$ for an antiferromagnet and
prefers the AFM order parameter to be aligned perpendicular to the
field ($g>0$).\cite{and80} Thus, for a general orientation of the field in the
basal plane there is competition between the anisotropy energy and the
Zeeman energy. The field at which the anisotropy energy is comparable
to the Zeeman energy defines a cross-over field scale,
\begin{equation}
H_{anis}=\sqrt{\frac{c'}{g}}\,m_0^2
\,.
\end{equation}
At low fields, $H \ll H_{anis}$, the AFM order parameter is essentially
locked by the in-plane anisotropy energy to an equilibrium orientation,
$\vartheta_n$. However, for $H > H_{anis}$ the Zeeman energy
dominates the anisotropy energy and the orientation of the
magnetization will adjust to remain approximately perpendicular to
$\vec{H}$. This case is particularly relevant to the discussion of
the hexagonal oscillations of $H_{c2}$. In the limit $H \gg H_{anis}$
the orientation of $\vec{m}_s$ is fixed by the field,
$\vartheta=\vartheta_H - \pi/2$, or $\vec{m}_s = m_s (\sin\vartheta_H,
-\cos\vartheta_H)$. The magnitude of the staggered moment is then to a very
good approximation determined by minimizing the AFM free energy
\begin{equation}
{\cal F}_{afm} = a(T)\, m_s^2 + b\, m_s^4 - c'\, m_s^6\,\cos(6\vartheta_H)
\,,
\end{equation}
at {\it fixed} orientation $\vartheta=\vartheta_H-\pi/2$. Treating the
anisotropy
energy as a perturbation leads to a small correction to the
exchange approximation for the AFM order parameter; $m_s = m_0 + \delta
m_s$. Retaining the leading order corrections to the stationarity
condition gives a hexagonal modulation of the AFM order parameter,
\begin{equation}
\frac{\delta m_s}{m_0} = \frac{3}{4}\left(\frac{U_{anis}}{U_{exch}}\right)\,
			\cos(6\vartheta_H)
\,,\quad H\gg H_{anis}\,.
\end{equation}
This modulation of the AFM order parameter by the anisotropy potential
leads to a hexagonal modulation of the upper
critical field in the SBF model for the double superconducting
transition.

\subsection{Coupling between Superconductivity and AFM}

I assume a specific 2D model, based on an E$_{2}$ order parameter,
which I discussed recently in the context of the H-T phase
diagram.\cite{sau94} In this model the gradient coefficients
$\kappa_2=\kappa_3\simeq 0$ for weak hexagonal anisotropy;\footnote{In
this context the term `weak hexagonal anisotropy' refers to the Fermi
surface. This anisotropy is the origin of the hexagonal anisotropy of
$H_{c2}^{\perp}(T)$ in perfect hexagonal systems, which vanishes for
$T\rightarrow T_c$ as $(1-T/T_c)^3$. This same Fermi surface anisotropy
is responsible for the deviation of $\kappa_{2,3}$ from zero in the
$E_2$ representations.\cite{sau94}} however, for in-plane magnetic
fields the principal arguments that follow apply to any of the 2D models. The
parameters of the GL functional in eq. \ref{f_sc} are calculated from
Fermi-liquid theory.\cite{muz93,sau94a} The gradient coefficients,
$\kappa_1\simeq N_f(v_f^{\perp})^2/(\pi T_c)^2$ and $\kappa_4\simeq
N_f(v_f^{||})^2/(\pi T_c)^2$, are determined by the density of states,
$N_f$, and the anisotropic components of the Fermi velocity,
$v_f^{\perp,||}$. These  coefficients determine the uniaxial anisotropy
of $H_{c1}$ and $H_{c2}$. The other coefficients,
$\alpha(T)=N_f(T-T_c)$ and $\beta_2=\frac{1}{2}\beta_1=\frac{7\zeta(3)
N_f}{16\pi^2 T_c^2}$, determine the condensation energy; $T_c$ is the
transition temperature in absence of coupling to the AFM order
parameter.

To calculate the hexagonal anisotropy of $H_{c2}$ near $T_c$ I need the
the terms in the free energy that describe the coupling of the
superconductivity to the AFM order parameter. The leading order
symmetry breaking terms are\cite{hes89,mac89,sau94}
\begin{equation}\label{f_sc-afm}
\begin{array}{l}
{\cal F}_{sc-afm} =
	-\epsilon\,m_s^2\,N_f T_c\,(|\eta_1|^2 - |\eta_2|^2) \\ \hspace{3em}
	+ \epsilon_{\perp}\,m_s^2\,\kappa_1\,
	(|\vec{D}_{\perp}\eta_1|^2 - |\vec{D}_{\perp}\eta_2|^2) \\  \hspace{3em}
	+ \epsilon_{||}\,m_s^2\,\kappa_4(|D_z\eta_1|^2 - |D_z\eta_2|^2)
\,.
\end{array}
\end{equation}
The first term is responsible for the double transition in zero field,
while the gradient terms generate an asymmetry in the slopes of the two
branches of $H_{c2}$ for $T>T^{*}$ and $T<T^{*}$.

Combining the coupling terms with the
superconducting free energy (eq. \ref{f_sc}) gives,
\begin{equation}
\begin{array}{l}
{\cal F}_{sc}+ {\cal F}_{sc-afm} =
	\alpha_{+}(T)|\eta_1|^2 + \alpha_{-}(T)|\eta_2|^2
\\ \hspace{6em}
	+ \kappa_{1}^{+}(\vec{D}_{\perp}\eta_1)\cdot(\vec{D}_{\perp}\eta_1)^{*}
	+ \kappa_{1}^{-}(\vec{D}_{\perp}\eta_2)\cdot(\vec{D}_{\perp}\eta_2)^{*}
\\ \hspace{6em}	+ \kappa_{4}^{+}(D_z\eta_1)(D_z\eta_1)^{*}
		+ \kappa_{4}^{-}(D_z\eta_2)(D_z\eta_2)^{*}
\,,
\end{array}
\end{equation}
with $\alpha_{\pm}(T) = N_f (T-T_c^{\pm})$. The fourth-order terms are
omitted since they are not relevant to the determination of $H_{c2}$.
The SBF field generates a splitting of $T_c$ and the gradient
coefficients,
\begin{equation}
T_{c}^{\pm} = T_{c}\,(1 \pm \epsilon m_s^2)
\qquad\,;\,\qquad
\kappa_{1,4}^{\pm} = \kappa_{1,4}\,
(1 \pm \epsilon_{\perp,||}\, m_s^2)
\,.
\end{equation}
The upper critical field is obtained from solutions of the linearized
GL equations,
\begin{equation}
\begin{array}{l}
\kappa_{1}^{+}\,\vec{D}_{\perp}^2\,\eta_{1}
	+ \kappa_{4}^{+}\,D_{z}^2\,\eta_{1}
	= \alpha_{+}(T)\,\eta_{1}
\,, \\
\kappa_{1}^{-}\,\vec{D}_{\perp}^2\,\eta_{2}
	+ \kappa_{4}^{-}\,D_{z}^2\,\eta_{2}
	= \alpha_{-}(T)\,\eta_{2}
\,.
\end{array}
\end{equation}
The resulting value of $H_{c2}^{\perp}(T)$ is then the larger value of the
solutions for the two branches,
\begin{equation}
H_{c2}^{\pm}(T;\vartheta_H)=\frac{\hbar c}{2e}\,
	\frac{N_f\,(T_{c}^{\pm}(\vartheta_H) - T)}
	     {\sqrt{\kappa_1^{\pm}\kappa_4^{\pm}}}
\,.
\end{equation}
For $\kappa_1^{+}\kappa_4^{+} > \kappa_1^{-}\kappa_4^{-}$ the upper
critical field has a kink at the temperature $T^{*}$ (tetracritical
point). The solution $H_{c2}^{+}$ denotes the branch with $T>T^{*}$ and
$H_{c2}^{-}$ denotes the branch with $T<T^{*}$. The two branches
correspond to transitions from the normal state to a superconducting
state with order parameters, $\vec{\eta}\sim(1,0)$ for $T>T^{*}$ and
$\vec{\eta}\sim(0,1)$ for $T<T^{*}$. Note that $H_{c2}^{\pm}(T)$ depends
implicitly on the orientation of the field, $\vartheta_H$, through the
AFM order parameter $m_s$.

Hexagonal oscillations of $H_{c2}^{\perp}(T)$ result from the
modulation of the AFM order parameter by the sixth-order anisotropy
energy. In the limit $H\gg H_{anis}$, the extrapolation points,
$T_{c}^{\pm}(\vartheta_H)$, show a hexagonal modulation. The modulation
can be scaled in units of the zero-field splitting of the transition,
$\Delta T_c /T_c = \epsilon\,m_0^2\,(1+\beta_1/\beta_2)$,
and the ratio of the anisotropy energy to the exchange energy,
\begin{equation}
T_{c}^{\pm}(\vartheta_H) = T_{c}^{\pm} \pm \frac{3}{2}\,
		\frac{\Delta T_c}{1+\beta_1/\beta_2}\,
		\left(\frac{U_{anis}}{U_{exch}}\right)\,
		\left[\cos(6\vartheta_H) - 1 \right]
\,.
\end{equation}
The parameters $T_c^{\pm}$ (without arguments) denote the
exptrapolation points in zero field.  Note that the modulation of the
two extrapolation points, $T_{c}^{+}$ and $T_{c}^{-}$, is out of phase
by $180^{o}$. This explains the change in phase of the oscillations of
$H_{c2}^{\perp}(T)$ for temperatures above and below the tetracritical
point. Figure 1 shows the upper critical field for two orientations of
the magnetic field.  Note the change in
$H_{c2}^{\perp}(T)$ upon rotating the field from $\vartheta_H = 0$
(minimum of the anisotropy energy) to $\vartheta_H = \pi/6$ (maximum of
the anisotropy energy). This change in phase of the oscillations for $T
> T^{*}$ and $T < T^{*}$ has the same origin as the splitting of $T_c$
in zero field by the AFM symmetry breaking field.\footnote{Note that
Fig. 1 includes the additional corrections discussed in the next
section; however, the dominant effect is the modulation of the
extrapolation temperatures by the anisotropy energy.}

The oscillations are small in magnitude,
\begin{equation}
|\delta H_{c2}^{\pm}|_{max} = \case{1}{2}
\left| H_{c2}^{\pm}(\vartheta_H = 0) - H_{c2}^{\pm}(\vartheta_H = \pi/6)\right|
\simeq\left|\frac{dH_{c2}^{\pm}}{dT}\right|
\,*\,\frac{3}{2}\,\frac{\Delta T_c}{(1+\beta_1/\beta_2)}\,
\left(\frac{U_{anis}}{U_{exch}}\right)
\,,
\end{equation}
and weakly dependent on temperature, except for $T\approx T^{*}$, where
the oscillations change sign (see Fig. 1), and for temperatures {\it
very} close to $T_c$ where $H_{c2}(T)$ drops below the anisotropy
field, $H_{anis}$. From the magnitude of the hexagonal anisotropy
measured by Keller, {\it et al.}\cite{kel94a}, $|\delta
H_{c2}|_{max}\simeq 10^{-3}\,{\rm T}$ for $T\simeq 507\,{\rm mK} > T^{*}\simeq
430\,{\rm mK}$, one obtains an estimate for the ratio of the anisotropy
energy to the exchange energy. Taking $\Delta T_c\simeq 60\,{\rm mK}$,
$dH_{c2}^{+}/dT\simeq 4.5\times 10^{-3}\,{\rm T/mK}$,\cite{kel94a} gives
$U_{anis}/U_{exch}\simeq 0.015$.

In addition to the modulation of the extrapolation temperatures, the
slopes of the two branches of $H_{c2}(T)$, are modulated by the
anisotropy potential,
\begin{equation}
S^{\pm}(\vartheta_H)=-dH_{c2}^{\pm}/dT =
	S^{\pm}\left(1 \mp (\epsilon_{||}+\epsilon_{\perp})\,m_0^2\,
	\frac{3}{4}\left(\frac{U_{anis}}{U_{exch}}\right)\,
	\left[\cos(6\vartheta_H) - 1\right] \right)
\,,
\end{equation}
where $S^{\pm}$ is the slope corresponding to a minimum of the
anisotropy energy. These terms generate temperature dependent
amplitudes for the oscillations of the two branches of
$H_{c2}^{\perp}(T)$ as shown in Fig. 1.

\subsection{Higher-order corrections at lower fields}

The oscillations of $H_{c2}^{\perp}(T)$ obtained above are for fields
large compared to the anisotropy field, $H_{anis}$. While it is
reasonable to assume that $H_{anis}$ is very low, it is non-zero and so
provides a cross-over field below which the hexagonal oscillations of
$H_{c2}^{\perp}$ should vanish. For $H > H_{anis}$ there are also
corrections to the `high-field' limit discussed above. These
corrections, as well as the low-field cross-over behavior, are obtained
from the minimization of the AFM free energy functional with respect to
both $m_s$ and $\vartheta$ for fixed $\vartheta_H$. The stationarity
condition $\delta{\cal F}_{afm}/\delta m_s = 0$ generates,
\begin{equation}\label{euler_ms}
\frac{\delta m_s}{m_0} = -\frac{3}{4}\left(\frac{U_{anis}}{U_{exch}}\right)
	\left\{ \cos(6\vartheta) + \frac{1}{3}\left(\frac{H}{H_{anis}}\right)^2\,
	\cos^2(\vartheta_H - \vartheta) \right\}
\,,
\end{equation}
to leading order in $U_{anis}/U_{exch}$. The second term in eq.
\ref{euler_ms} originates from the Zeeman energy. The relation between
the orientation of the applied field ($\vartheta_H$) and that of the
AFM order parameter ($\vartheta$) is determined by the stationarity
condition $\delta{\cal F}_{afm}/\delta \vartheta = 0$,
\begin{equation}\label{euler_theta}
\sin(6\vartheta) =
\frac{1}{6}\left(\frac{H}{H_{anis}}\right)^2\,\sin(2(\vartheta_H - \vartheta))
\,.
\end{equation}
In the zero-field limit one obtains the expected
result that the AFM order parameter is locked to the lattice at the
minima of the anisotropy energy given by $n\pi/3 + \pi/6$ for $c' > 0$;
the solutions $n\pi/3$ correspond to maxima of the anisotropy energy.
In the opposite limit, $H_{anis}/H \rightarrow 0$, the AFM order
parameter is oriented at $\pm\pi/2$ relative to the field; the other
solutions at $\pm \pi$ corrrespond to maxima of the Zeeman energy.

For $H_{anis}/H \ll 1$ the orientation of $\vec{m}_s$ is no longer
strictly orthogonal to $\vec{H}$. The deviation, $\beta = (\vartheta_H
- \vartheta) - \pi/2$, oscillates  with an amplitude proportional to
$(H_{anis}/H)^2$,
\begin{equation}
\beta \simeq 3\left(\frac{H_{anis}}{H}\right)^2\,\sin(6\vartheta_H)
\,.
\end{equation}
This phase shift vanishes near a minimum of the anisotropy energy and
is maximum at orientations of $\vec{H}$ corresponding to a maximum in
the slope of the anisotropy energy ({\it i.e.} maximum anisotropy
torque on $\vec{m}_s$). This correction, combined with the correction
from the Zeeman energy, gives
\begin{equation}\label{delta_ms}
\frac{\delta m_s}{m_0} = \frac{3}{4}\left(\frac{U_{anis}}{U_{exch}}\right)
	\, f(\vartheta_H)
\,,
\end{equation}

\begin{equation}
f(\vartheta_H) =
\cos\left(6\vartheta_H -
18\left(\frac{H_{anis}}{H}\right)^2\sin(6\vartheta_H)\right) -
3\left(\frac{H_{anis}}{H}\right)^2\,
\sin^2(6\vartheta_H)
\,,
\end{equation}
for the modulation of the AFM order parameter parameter. The
corrections from the Zeeman energy and the angular deviation from
$\pi/2$ generate non-sinusoidal hexagonal oscillations of $H_{c2}(T)$.
Figure 2 shows these oscillations for temperatures above and below
$T^{*}$. Note the sharpening of the oscillations for $\vec{H} ||
\vec{a}^{*}$ ($\vartheta_H=\pi/6$). A similar effect is also seen in the
experimental results for $H_{c2}^{\perp}(T)$ reported by Keller, {\it
et al.}\cite{kel94a}

At very low fields, $H < H_{anis}$, and therefore at temperatures very
near $T_c$ ({\it i.e.} $|T-T_c| < H_{anis}/|dH_{c2}^{+}/dT|$), the
hexagonal oscillations of $H_{c2}^{\perp}$ disappear. The AFM order
parameter becomes locked to the lattice by the in-plane anisotropy
energy as $H\rightarrow 0$. For $H\ll H_{anis}$, the Zeeman energy is a
perturbation and leads to small deviations of $\vec{m}_s$ from the
minima of the anisotropy energy. Assuming that $\vec{m}_s$ is oriented
at $\vartheta = \pi/6$ in zero field (corresponding to a minimum along
the $\vec{a}^{*}$ axis), the perturbed orientation of $\vec{m}_s$ at
finite $\vec{H}$ is,
\begin{equation}
\vartheta = \pi/6 - \frac{1}{36} \left(\frac{H}{H_{anis}}\right)^2\,
		\sin(2\vartheta_H - \pi/3)
\,.
\end{equation}
This deviation leads to an {\it orthorhombic} modulation of the AFM order
parameter,
\begin{equation}
\frac{\delta m_s}{m_0} = \frac{3}{4}\left(\frac{U_{anis}}{U_{exch}}\right)\,
	\left[ 1 - \frac{1}{3} \left(\frac{H}{H_{anis}}\right)^2
	\cos^2(\vartheta_H - \pi/6) \right]\,,
	\quad H \ll H_{anis}\,
\,.
\end{equation}
For $H_{c2}(T)\ll H_{anis}$ and $\vartheta_H - \pi/6 \ne \pm\pi/2$ the upper
critical field line
bends towards the zero-field transition temperature, $T_{c+}$,
\begin{equation}\label{crossover}
H_{c2}(T) = S^{+}\,\left[ (T_{c}^{+} - T) -
	\frac{3}{2}\left(\frac{\Delta T_c}{1+\beta_1/\beta_2}\right)
	\left(\frac{U_{anis}}{U_{exch}}\right)
	\left(\frac{S^{+} (T_{c}^{+} - T)}{H_{anis}}\right)^2\,
	\cos^2(\vartheta_H - \pi/6) \right]
\,.
\end{equation}

In conclusion, I have developed a Ginzburg-Landau theory for the upper
critical field anisotropy of UPt$_3$ based on a superconducting order
parameter belonging to a two-dimensional representation coupled to an
in-plane AFM order parameter. The key terms responsible for the
hexagonal anisotropy of $H_{c2}$ near $T_c$ are the in-plane anisotropy
and Zeeman energies of the AFM phase. These terms lead to hexagonal
modulations of the AFM order parameter, which generate hexagonal
anisotropy in $H_{c2}(T)$ through the SBF coupling to
superconductivity. The GL theory accounts for the basic features of the
observed oscillations, including the sign reversal at $T\approx T^*$.

\section{Acknowledgements}

I acknowledge the support of the Aspen Center for Physics where this
work developed, and the NSF through grant no. 9120521 for the
Materials Research Center at Northwestern University. I thank V. P.
Mineev and M. Zhitomirsky for their comments and critique, and
especially M. Norman for important conversations. I also thank N.
Keller for discussions and for sending me a copy of the work reported
in Refs. \onlinecite{kel94,kel94a} prior to publication.

\newpage

\begin{figure}
\centerline{\psfig{figure=Hc2.ps,width=4.5in}}
\begin{quote}
\small
Fig. 1{\hskip 10pt} Anisotropy of $H_{c2}^{\perp}(T)$ vs. $T/T_c$. The
solid (dashed) curve is $H_{c2}^{\perp}(T)$ for $\vartheta_H = 0$
[$\vec{H}||\vec{a}$] ($\vartheta_H = \pi/6$ [$\vec{H}||\vec{a}^*$]).
The cross-over occurs in the vicinity of $T^*\simeq 0.8 T_c$. The
magnitude of the anisotropy and the width of the transition region are
amplified by the large value of $U_{anis}/U_{exch} = 0.2$ (chosen to
show the anisotropy clearly in the graph). The other parameters are
$S^-/S^+ =1.5$, $\epsilon m_0^2 = 0.03$, $(\epsilon_{||} +
\epsilon_{\perp}) m_0^2 = 0.2$, $H_{anis}/S^+ T_c = 0.04$. Below
$H_{anis}$ (dotted line) the curves merge at $T_c$ (sketch).

\end{quote}
\end{figure}

\medskip

\begin{figure}
\centerline{\psfig{figure=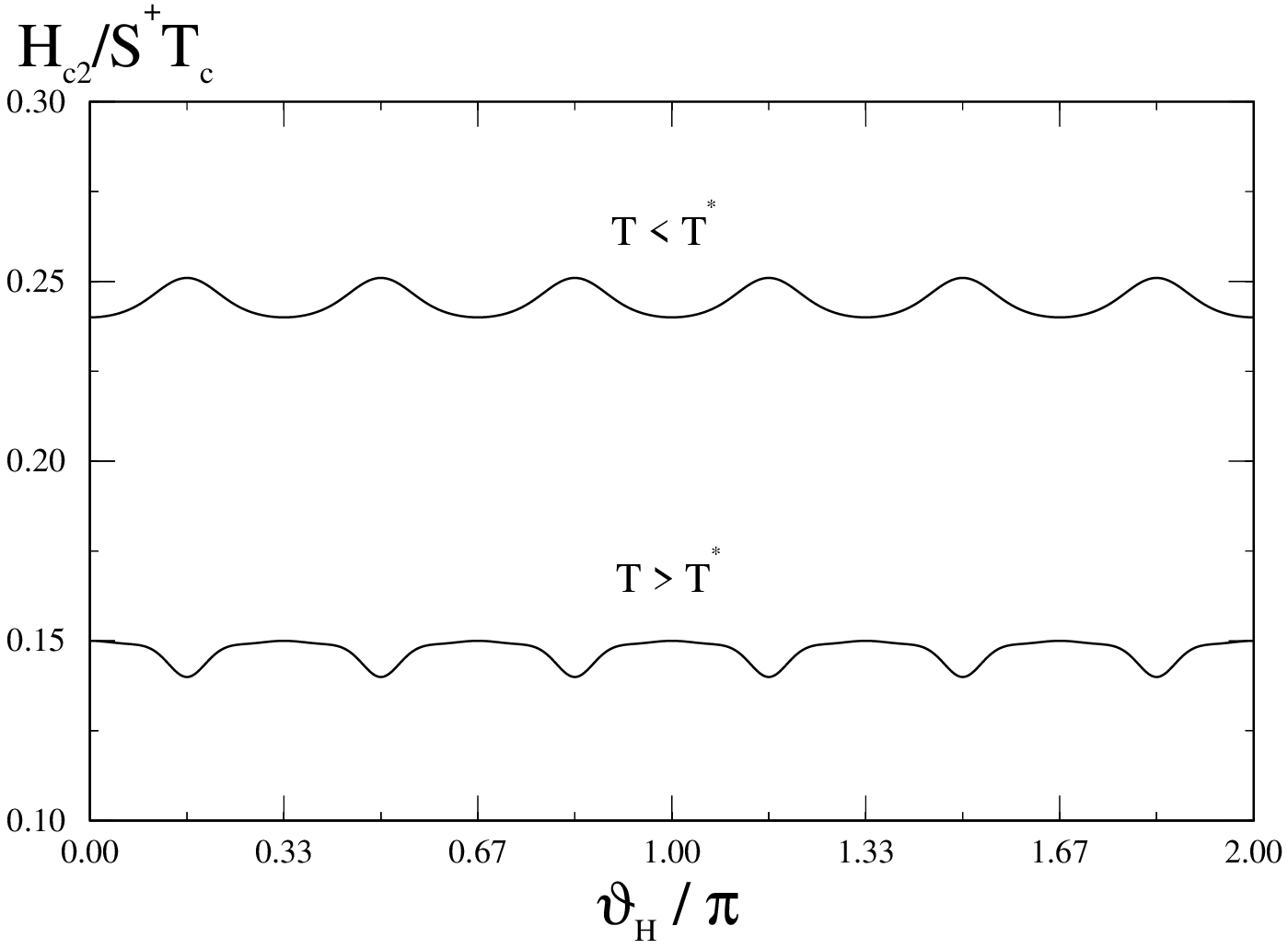,width=4.5in}}
\begin{quote}
\small
Fig. 2{\hskip 10pt} Anisotropy of $H_{c2}^{\perp}(T)$ vs. $\vartheta_H$.
The angular dependence of $H_{c2}$ shows the $180\,^o$ relative phase
of the oscillations for $T>T^*$ and $T<T^*$. The parameters are the
same as those in Fig. 1. The scalloped shape of the oscillations arises
from the deviation of $\vartheta_H - \vartheta$ from $\pi/2$.

\end{quote}
\end{figure}

\end{document}